\documentclass[twoside]{article}
\usepackage{graphicx} 
\usepackage[natbibapa]{apacite}
\usepackage{colortbl}
\usepackage{anyfontsize}

\usepackage{subcaption}  
\usepackage{authblk}

\usepackage{float}

\newfloat{textbox}{tbhp}{lop}
\floatname{textbox}{Box}

\usepackage{multirow}
\usepackage{multicol}
\usepackage{array}
\usepackage{xcolor}
\usepackage{layout}

\usepackage[framemethod=default]{mdframed}
\usepackage[top=1.25in, bottom=1.25in,
            left=1in, right=1in]{geometry}
\usepackage{tcolorbox}
\usepackage{caption}

\usepackage[toc,page]{appendix} 
\newtcolorbox{textexcerpt}[2][]{%
  colback=white,
  colframe=black,
  fonttitle=\bfseries,
  #1
}

\usepackage{orcidlink}

\usepackage{authblk}

\usepackage{amsmath}

\usepackage{float}

\usepackage{fancyhdr}

\usepackage{listings}
\definecolor{codegreen}{rgb}{0,0.6,0}
\definecolor{codegray}{rgb}{0.5,0.5,0.5}
\definecolor{codepurple}{rgb}{0.58,0,0.82}
\definecolor{backcolour}{rgb}{0.95,0.95,0.92}

\lstdefinestyle{mystyle}{
    backgroundcolor=\color{backcolour},   
    commentstyle=\color{codegreen},
    keywordstyle=\color{magenta},
    numberstyle=\tiny\color{codegray},
    stringstyle=\color{codepurple},
    basicstyle=\ttfamily\footnotesize,
    breakatwhitespace=false,         
    breaklines=true,                 
    captionpos=b,                    
    keepspaces=true,                 
    numbers=left,                    
    numbersep=5pt,                  
    showspaces=false,                
    showstringspaces=false,
    showtabs=false,                  
    tabsize=2
}

\title{Deep Learning Model Deployment in Multiple Cloud Providers: an Exploratory Study Using Low Computing Power Environments}

\author[1]{Elayne Lemos\orcidlink{0000-0002-7347-8692}}
\author[2]{Rodrigo Oliveira}
\author[3]{Jairson Rodrigues\orcidlink{0000-0003-1176-3903}}
\author[3]{Rosalvo F. Oliveira Neto\orcidlink{0000-0002-3290-5539}\thanks{CONTACT Rosalvo F. Oliveira Neto. Email: rosalvo.oliveira@univasf.edu.br}}

\affil[1]{Cloud Plarform Engineering, Wildlife Studios, Dr. Renato Paes de Barros St, 1017, 11 floor - Itaim Bibi, São Paulo-SP, Brazil}
\affil[2]{Technology, State University of Feira de Santana, Bahia, Brazil}
\affil[3]{Computer Engineering, Federal University of Vale do São Francisco, Bahia, Brazil}

\begin{document}

\pagestyle{fancy}
\fancyhf{}
\fancyhead[RE]{\textit{LEMOS et al.}}
\fancyhead[LO]{\textit{Deep Learning Model Deployment in Multiple Cloud Providers}}
\fancyfoot[C]{\thepage}

\renewcommand\thefootnote{}
\footnotetext{\textbf{Abbreviations:} ML, machine learning; POC, proof of concept; DL, deep learning}

\renewcommand\thefootnote{\fnsymbol{footnote}}
\setcounter{footnote}{1}

\maketitle

\begin{abstract}

\noindent The deployment of Machine Learning models in the cloud has grown among tech companies. Hardware requirements are higher when these models involve Deep Learning techniques, and the cloud providers' costs may be a barrier. We explore deploying Deep Learning models, using for experiments the GECToR model, a Deep Learning solution for Grammatical Error Correction, across three of the major cloud providers (Amazon Web Services, Google Cloud Platform, and Microsoft Azure). We evaluate real-time latency, hardware usage, and cost at each cloud provider in 7 execution environments with 10 experiments reproduced. We found that while Graphics Processing Units (GPUs) excel in performance, they had an average cost 300\% higher than solutions without a GPU. Our analysis also suggests that processor cache memory size is a key variable for CPU-only deployments, and setups with sufficient cache achieved a 50\% cost reduction compared to GPU-based deployments. This study indicates the feasibility and affordability of cloud-based Deep Learning inference solutions without a GPU, benefiting resource-constrained users such as startups and small research groups.

\end{abstract}


\section{Introduction}\label{sec:intro}

Classical Machine Learning (ML) is known for its reliance on manual feature engineering, a process that often requires extensive domain knowledge and significant effort in data preprocessing and transformation \citep{Neto2017}. To achieve a more scalable scenario in Artificial Intelligence, there is a desire to make algorithms less dependent on feature engineering \citep{Bengio2014}. The solution to this demand came in the form of Deep Learning algorithms, which allow computational models composed of multiple processing layers to learn data representations with multiple levels of abstraction. Deep Learning (DL) is a subfield of ML that encompasses representation learning. The latter refers to a set of methods that allows a machine to automatically identify the necessary representations for detection or classification from raw data, i.e., automating feature engineering \citep{Lecun2015}.

Recent breakthroughs in Natural Language Processing (NLP) are based on large DL models such as BERT and GPT-2 \citep{Gonzalez2020,Wolf2020}. However, in developing countries, the cost of training these large models can be prohibitive due to the high prices of hardware with high computational power requirements, such as the GPU \citep{Mittal2019}. An approach to overcome this issue is to adopt cloud computing services to save money by using the resources on-demand instead of purchasing expensive hardware. Businesses can use services in the cloud and pay temporary access fees. Although cloud computing reduces the cost in a model’s training phase, a question arises: What is the viability for a startup in developing countries of running a Proof of Concept (POC) with these models in a cloud environment? A POC is a strategy used to determine whether an innovation or product is feasible \citep{Mittal2019}. In general, conducting a POC consists of deploying a solution in a production environment without redundancy and high availability requirements. This paper aims to analyze the feasibility of carrying out POCs in the cloud. For this, a case study was conducted by applying Grammatical Error Correction (GEC) on Amazon Web Services (AWS), Google Cloud Platform (GCP), and Microsoft Azure.

The remainder of this paper is organized as follows. Section \ref{sec:review} lists related works and provides a brief presentation of Machine Learning as a Service (MLaaS) and GEC. Section \ref{sec:method} discusses the experimental methodology used in the research. Section \ref{sec:results} presents the results and their interpretation. Finally, Sections \ref{sec:conclusion} and \ref{sec:future} conclude this paper and discuss limitations with recommendations.

\subsection{Objectives}\label{sec:objectives}

This research aims to evaluate the feasibility of deploying DL models on low-cost cloud-based MLaaS platforms. The performance of these models under varying loads was analyzed by simulating simultaneous queries. The analysis focused on key metrics such as response time, memory and vCPU consumption, and infrastructure cost. By deploying the MLaaS solution on different hardware configurations, the most cost-effective options for running complex DL models in the cloud were identified.

\subsection{Justification}

\begin{figure}[ht]%
\centering
\includegraphics[width=0.85\textwidth]{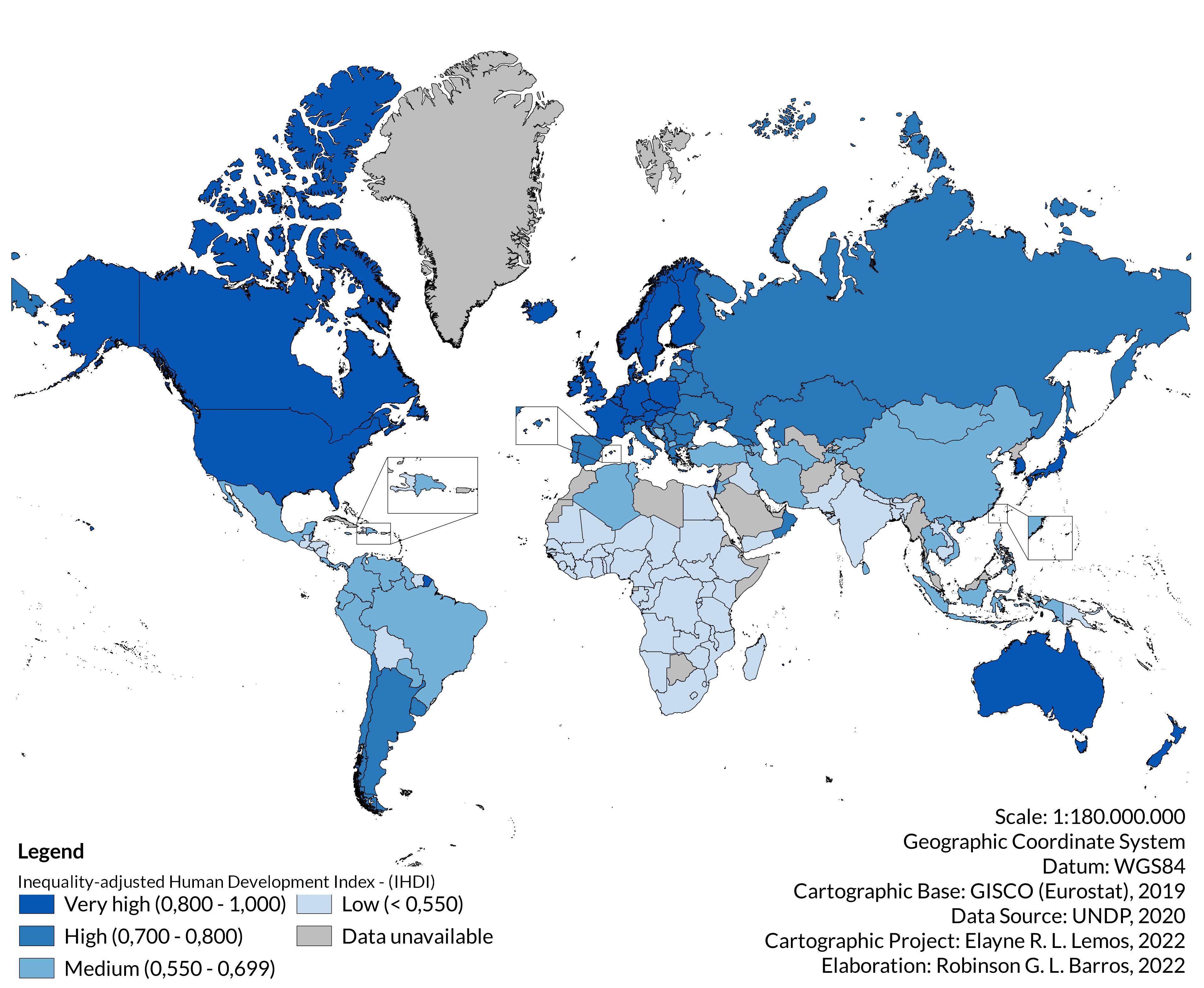}
\caption{\label{img:idhad} Inequality-adjusted Human Development Index by country.}
\end{figure}

\begin{figure}[ht]%
\centering
\includegraphics[width=0.85\textwidth]{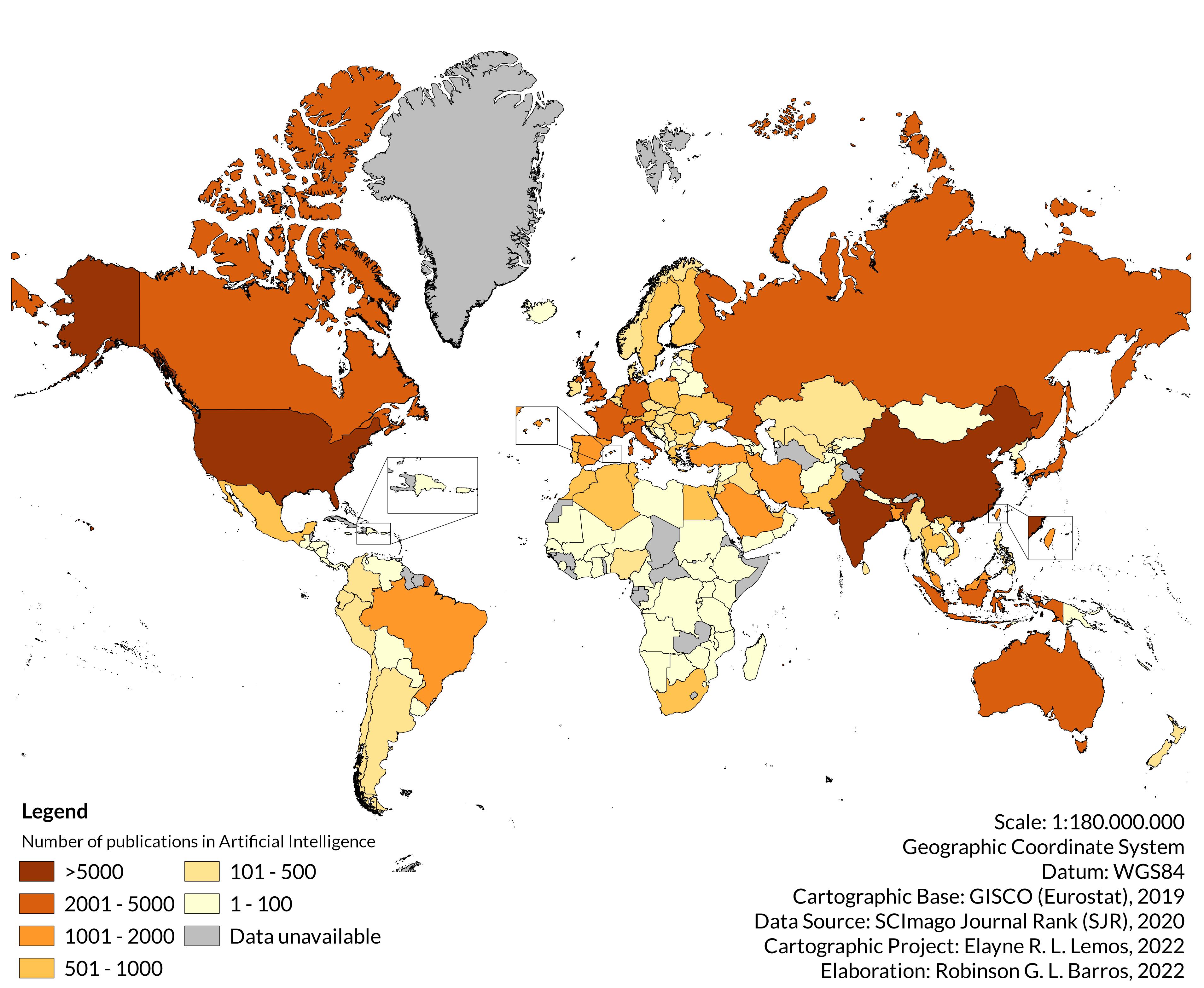}
\caption{\label{img:pubs} Artificial Intelligence publications by country.}
\end{figure}

The global digital divide, characterized by unequal access to technological resources, remains a serious hurdle \citep{ILO2019,UNDP2020}. This lack of access particularly hinders research and innovation in Artificial Intelligence (AI), a field requiring substantial computational power and specialized infrastructure. Figure \ref{img:idhad} illustrates the stark disparities in the Inequality-Adjusted Human Development Index across territories. These inequities translate into limitations for developing regions to participate in cutting-edge AI research, potentially hindering their ability to leverage AI solutions for crucial areas like healthcare, education, and economic growth. Furthermore, there is a strong correlation between a country's level of development and its contribution to the field of AI, as evidenced by the distribution of scientific publications shown in Figure \ref{img:pubs}.

\subsection{Questions}\label{sec:questions}

Deploying ML models in production environments is crucial for real-world impact.  Cloud providers offer a convenient solution for hosting large-scale MLaaS offerings \citep{Singh2021}. However, deploying DL models presents an important challenge: their hardware requirements. Particularly, the need for specialized GPUs can significantly increase costs. This is especially concerning for developing countries, where cloud services are often priced in US dollars, adding a financial hurdle. To effectively deploy DL models in the cloud while staying within budget constraints, it's crucial to carefully consider the hardware configuration. The present work frames three key questions that will guide decision-making and optimize cloud deployment:

\begin{enumerate}
    \item How much RAM is required for acceptable performance?
    \item How many virtual CPUs (vCPUs) are necessary to handle the expected workload?
    \item Can the model achieve acceptable performance without a GPU?
\end{enumerate}

\begin{figure}[ht]%
\centering
\includegraphics[width=0.3\textwidth]{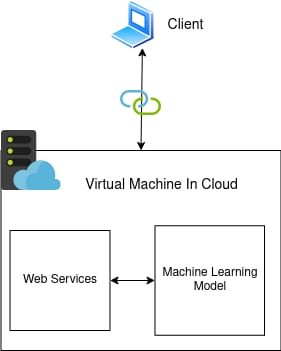}
\caption{\label{img:mlaas} MLaaS solution layers: client, web service, and ML model}
\end{figure}

Figure \ref{img:mlaas} illustrates an architecture in three logical layers of an MLaaS solution in the context of the POC to be developed. The first layer consists of the client submitting requests to the web service (second layer). The third layer corresponds to the query process of the ML model and the consequent response.

In the experiments, the logical layers of the web service and ML model will reside in a single physical network node, represented in Figure 3\ref{img:mlaas} by the virtual machine instance in the cloud. To contextualize this approach, the following section reviews relevant literature in the field.

\section{Literature Review}\label{sec:review}

\subsection{Cloud Computing and Machine Learning as a Service}

\begin{figure}[ht]%
\centering
\includegraphics[width=0.55\textwidth]{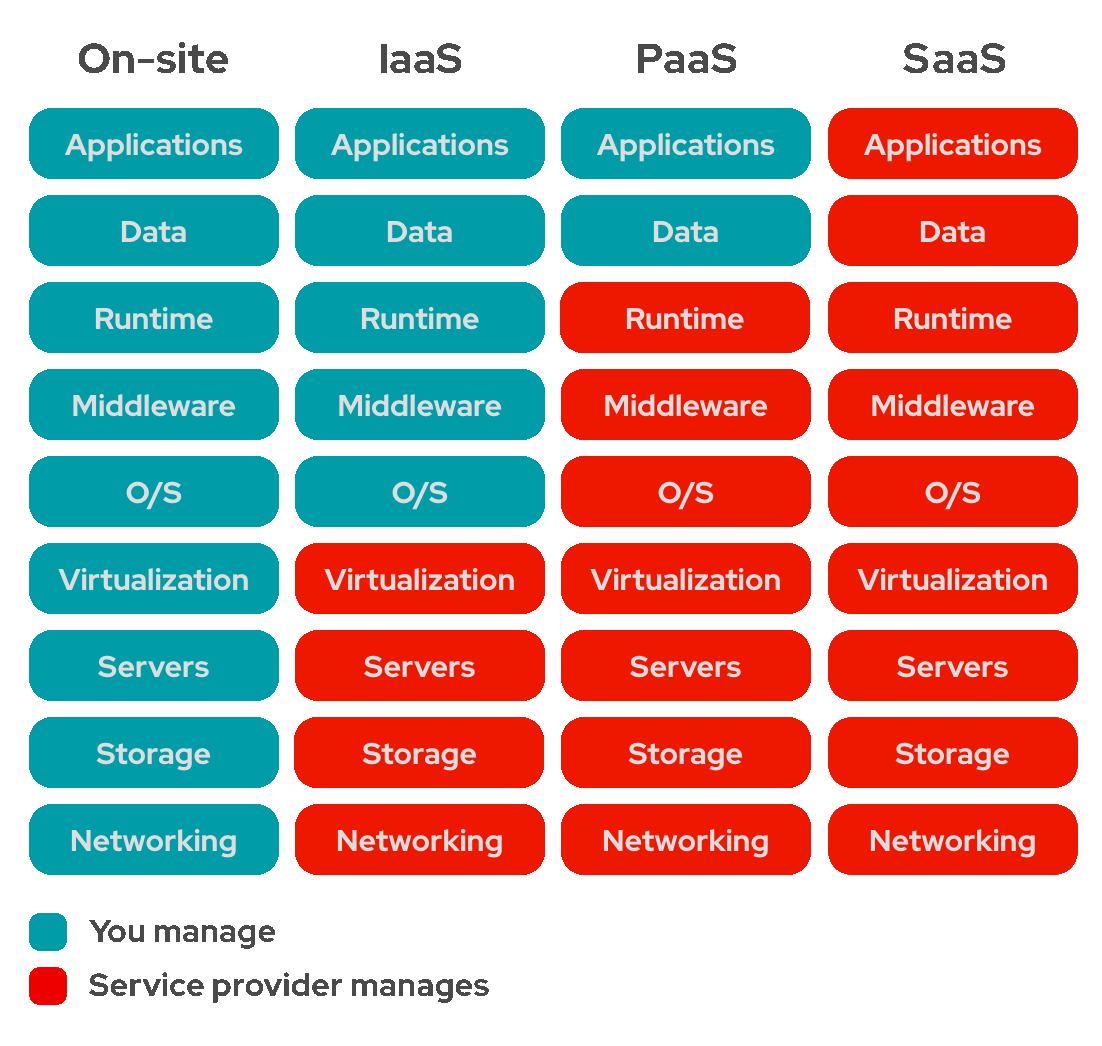}
\caption{\label{img:xaas} Computational application environments regarding the level of abstraction for the user. Source: \citet{RedHat2022}.}
\end{figure}

Cloud Computing represents a computational model for convenient, on-demand access from any location to servers, storage, applications, and services readily available with minimal management effort \citep{Mell2011}. Computational clouds can be conceived and implemented in the form of functionally distinct models, generically represented by the term Everything as a Service (XaaS), where X stands for software (SaaS), platform (PaaS), or Infrastructure (IaaS) \citep{Fox2009}. The Infrastructure as a Service (IaaS) model is quite common in the form of complete virtual machines with their respective virtualized hardware resources \citep{Rodrigues2020}. Figure \ref{img:xaas} graphically represents the abstraction levels of the different cloud computing business models, whose details can be seen below:

\begin{itemize}
    \item Infrastructure as a Service (IaaS) is the provision of computing resources, such as network, storage, or servers, by cloud providers through virtualization, enabling infrastructure scalability and avoiding the need for the user to keep these resources on their own hardware. Some of the so-called public cloud providers are AWS, Azure, and GCP.
    \item Platform as a Service (PaaS) offers an environment where users can develop, manage, and deliver applications in the cloud, abstracting the management of the infrastructure layer or the production environment in charge of the provider. A typical example is web server provisioning platforms such as Heroku.
    \item Software as a Service (SaaS) offers access to cloud-based software, not requiring installation on individual machines, commonly through a subscription model in which updates, maintenance, and corrections are the provider’s responsibility. It is considered the most comprehensive form of cloud computing services, and the user connects to the solution through a control panel, a web browser, or an API.
\end{itemize}

MLaaS refers to a model that comprises a set of cloud-based platforms that serve the development and deployment phases of ML solutions \citep{Fox2009,Ribeiro2015}. These platforms enable data management and access to ML tools, such as APIs and predictive models, in an infrastructure with high processing power, such as GPUs.

Among the leading platforms for MLaaS, we can mention: Amazon Machine Learning, Azure Machine Learning, Google AI Platform, and IBM Watson. These platforms charge per hour of use, which is a viable and sometimes cheaper alternative in model development since this way of pricing is more affordable than buying the hardware. However, for production environments where the service must be available 24 hours a day, this model may incur high costs, sometimes difficult to predict, making its use unfeasible.

\subsection{Grammatical Error Correction And Deep Learning}

\begin{figure}[ht]%
\centering
\includegraphics[width=0.55\textwidth]{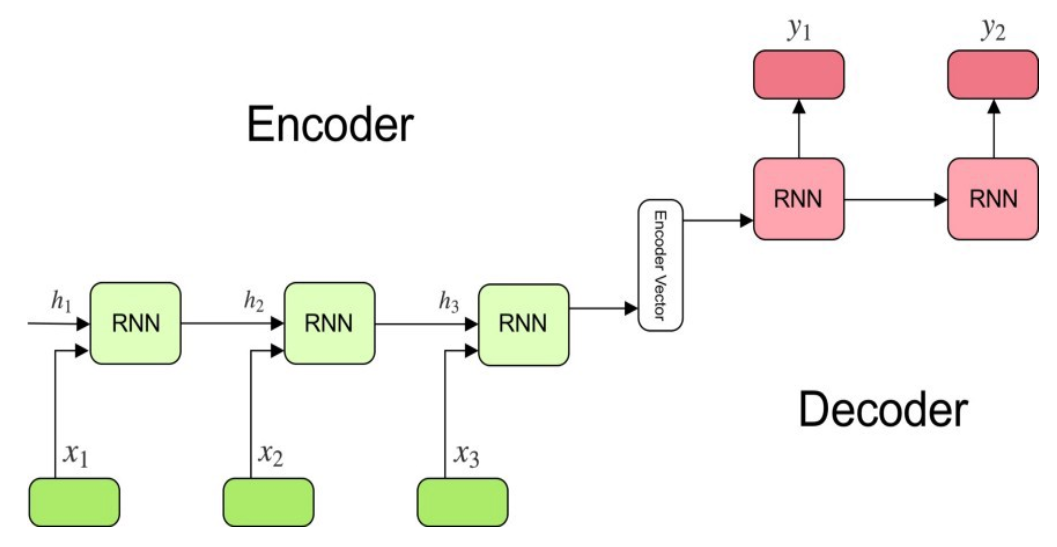}
\caption{\label{img:enc-dec} Encoder-decoder neural network architecture. Source: \citet{Kostadinov2019}.}
\end{figure}

GEC is an NLP task that aims to identify and automatically correct grammatical errors (syntactic or semantic) in written texts. A GEC solution receives a text, analyzes it in search of errors, and corrects them, keeping its original meaning \citep{Raheja2020}. There are several approaches to building a GEC solution. The most used approach today is the Neural Machine Translation (NMT) \citep{Yuan2016}. According to \citet{Ghader2017}, the core architecture of NMT models is based on the general encoder-decoder approach \citep{Sutskever2014}. Figure \ref{img:enc-dec} shows an end-to-end approach that learns to encode source sentences into distributed representations and decode these representations into sentences in the target language.

Since this is a DL architecture, it requires high processing time and memory consumption. In this study, we used the Grammarly pre-trained model \citep{Omelianchuk2020} with a 500 MB neural network file.

\subsection{Related works}

The relevance of the research on time and cost optimization is highlighted by research found in the academic literature in the last two decades, as evidenced below. In \citet{Rodrigues2020b,Rodrigues2021}, the authors investigate hardware and data volume parameters to discover the most relevant factors regarding time and cost in the execution of ML algorithms on big datasets in public cloud computing environments. The authors obtained linear regression models to estimate the time and cost of executing these algorithms based on the analyzed factors. It was possible to accurately quantify the impact of the number of machine nodes on time performance. Similar work was carried out to analyze the impact of software configuration parameters with a fixed hardware infrastructure, also in clusters hosted in public cloud environments \citep{Nunes2023}. The author obtained performance prediction models using backward stepwise linear regression. Although such works have found interesting results in their scopes, all of them used high-performance distributed cluster infrastructure. Thus, there is a gap in performance investigation considering environments with reduced computational power that employ Deep Learning techniques. It is still possible to highlight works focused on time performance optimization running on traditional (non-distributed) computing in tasks focused on Support Vector Machines (SVM) \citep{Staelin2003}, neural networks \citep{Packianather2000}, and evolutionary computing \citep{Bates2004,Ridge2007,Pais2014}. However, none of these works focused on solving Deep Learning problems. It is also common to find specific investigations on DL that perform benchmark tests comparing hardware architectures. In \citet{Wang2019}, for example, the authors systematically benchmark models for fully connected (FC), convolutional (CNN), and Recurrent Neural Networks (RNN) on Google’s Cloud TPU v2/v3, NVIDIA’s V100 GPU, and an Intel Skylake CPU platform. They quantify the rapid performance improvements that specialized software stacks provide for the TPU and GPU platforms. Again, despite the valuable contributions, there was no focus on low-resource machines in the cloud.

We have not found any work exclusively focused on evaluating real-time deep learning solutions in low-resource environments in the cloud. Therefore, this article presents itself as an exploratory investigation of performance in terms of monetary costs and execution time of deep learning models in public clouds using hardware with low computational power. Based on these gaps in the literature, the following section presents the proposed experimental design.

\section{Material and methods}\label{sec:method}

A load test was conducted to assess the feasibility of executing a deep learning POC in a resource-constrained environment (without a GPU) in the cloud using GECToR. The aim was to evaluate the solution's response time, vCPU, and memory loads under varying workload scenarios and computational configurations, gradually increasing demand in three different production environments. AWS, GCP, and Azure were selected as cloud infrastructure providers based on their cost-effectiveness and leadership position in the industry.

Five non-GPU hardware configurations were chosen in each of the cited cloud providers, endowed with the minimum memory requirements necessary for the execution of the GECToR model. The minimum memory configuration was based on exploratory investigations by the authors during the experimental period. It considered the size of the model (7 GB) to be loaded and the requirements of the Operating System (OS) and support services for the MLaaS solution (1 GB RAM). In addition, two more GPU configurations were selected for comparison. Table \ref{tab:instances} displays the information about the seven investigated machine configurations for all providers, totaling twenty-one scenarios.

\begin{table}%
\def\arraystretch{1.5}%
\fontsize{7pt}{7pt}\selectfont
\centering
\caption{\label{tab:instances} Machines’ configurations on AWS, GCP, and Azure cloud providers.}
\begin{tabular}{clccccc}
\multicolumn{4}{c}{\textbf{Legend:} (1) Machine label ~ ~ (2) Cache Memory   ~ ~ (3) Random Access Memory (GB)} \\
\hline
\textbf{Cloud} & \textbf{Instance / Physical Processor}     & \textbf{L}$^{(1)}$ & \textbf{vCPUs} & \textbf{C}$^{(2)}$ & \textbf{RAM}$^{(3)}$ & \textbf{GPU} \\ \hline
    & c6a.xlarge / AMD EPYC 7R13 (2.95 Ghz)                 & A          & 4              & 2          & 8            & -         \\
    & c6a.2xlarge / AMD EPYC 7R13 (2.95 Ghz)                & B          & 8              & 2          & 16           & -         \\
    & t2.xlarge / Intel Xeon Scalable (3.3 Ghz)             & C          & 4              & 4          & 16           & -         \\
AWS & inf1.xlarge / Intel Xeon Platinum 8275CL (3.0 GHz)    & D          & 4              & 2          & 8            & -         \\
    & inf1.2xlarge / Intel Xeon Platinum 8275CL (3.0 GHz)   & E          & 8              & 2          & 16           & -         \\
    & g4dn.xlarge / Intel Xeon Platinum 8259CL (2.5 GHz)    & F          & 4              & -          & 16           & NVIDIA T4 \\
    & g4dn.2xlarge / Intel Xeon Platinum 8259CL (2.5 GHz)   & G          & 8              & -          & 32           & NVIDIA T4 \\ \hline

    & n2d-custom-4-8192 / AMD EPYC Milan 7B13 (3.5 GHz)     & A          & 4              & 2          & 8            & -         \\
    & n2d-custom-8-16384 /AMD EPYC Milan 7B13 (3.5 GHz)     & B          & 8              & 2          & 16           & -         \\
    & n2-custom-8-16384 / Intel Xeon Gold 6268CL (3.9 GHz)  & C          & 4              & 4          & 16           & -         \\
GCP & c3-highcpu-4 / Intel Xeon Platinum 8481C (3.3 GHz)    & D          & 4              & 2          & 8            & -         \\
    & c3-highcpu-8 / Intel Xeon Platinum 8481C (3.3 GHz)    & E          & 8              & 2          & 16           & -         \\
    & n1-standard-4 / Intel Xeon Platinum 8173M (3.5 GHz)   & F          & 4              & -          & 16           & NVIDIA T4 \\
    & n1-standard-8 / Intel Xeon Platinum 8173M (3.5 GHz)   & G          & 8              & -          & 32           & NVIDIA T4 \\ \hline

      & standard\_B4als\_v2 / AMD EPYC Milan 7763v (3.5 GHz)    & A          & 4              & 2          & 8            & -         \\
      & standard\_B8als\_v2 / AMD EPYC Milan 7763v (3.5 GHz)    & B          & 8              & 2          & 16           & -         \\
      & standard\_D8lds\_v5 / Intel Xeon Platinum 8370C (3.5 GHz) & C          & 4              & 4          & 16           & -         \\
Azure & standard\_F4s\_v2 / Intel Xeon Platinum 8370C (3.7 GHz) & D          & 4              & 2          & 8            & -         \\
      & standard\_F8s\_v2 / Intel Xeon Platinum 8370C (3.7 GHz) & E          & 8              & 2          & 16           & -         \\
      & standard\_NC4as\_T4\_v3 / AMD EPYC Rome 7V12 (3.3 GHz)    & F          & 4              & -          & 28           & NVIDIA T4 \\
      & standard\_NC8as\_T4\_v3 / AMD EPYC Rome 7V12 (3.3 GHz)    & G          & 8              & -          & 56           & NVIDIA T4 \\ \hline
\end{tabular}
\end{table}

\subsection{Simulation Corpus}\label{sec:corpus}

The CoNLL-2014 database has become a benchmark in the evaluation of GEC solutions by incorporating the NUS Corpus of Learner English (NUCLE), considered revolutionary due to its size at the time it was launched as a large, fully annotated database available free for research. Thus, the present work uses the NUCLE 3.2 test base, the same one used in CoNLL-2014 for the proposed experiments \citep{Ng2014}. The test was composed of essays by 25 students from the National University of Singapore in response to two proposed questions, each essay containing an average of 500 words, with a total of 50 essays, with 1312 sentences and 30144 tokens. The frequency of grammatical errors in the database is considered low, which can be explained by the greater proficiency in English expected of university students \citep{Dahlmeier2013}.

\subsection{Pre-trained model}\label{sec:model}

There are pre-trained models in GEC with good results available for free. For the experiments, we used the open source model proposed by \citet{Omelianchuk2020}, due to its relevant result, reaching an F0.5 of 65.3\% in the CoNLL-2014 test corpus. This GEC system has an approach based on seq2seq, with an encoder composed of pre-trained BERT transformers \citep{Wolf2020,Gonzalez2020}, in their base configurations, stacked with two linear layers with a softmax layer on top to perform operations on the text. For the best configuration of GECToR, the estimated inference time in its development was 0.4s, using an Nvidia Tesla V100 GPU at a batch size of 128 in the infrastructure.

\subsection{Software architecture for simulation}\label{sec:architecture}

\begin{figure}[ht]%
\centering
\includegraphics[width=0.75\textwidth]{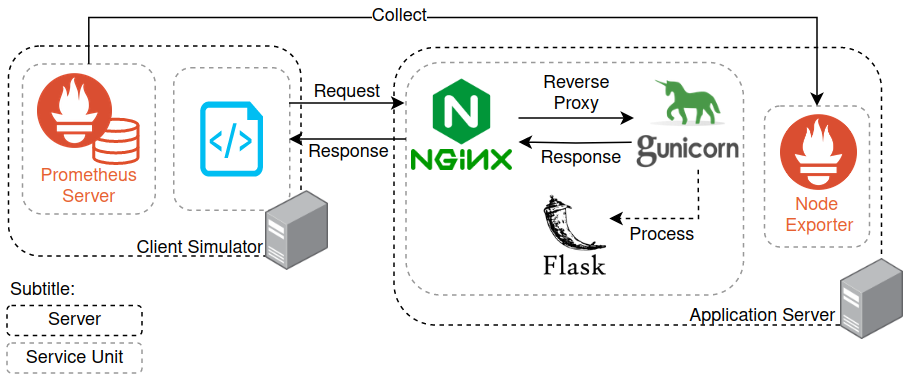}
\caption{\label{img:soft-arch} Simulation system architecture used for the experiments, with the client instance responsible for emulating the requests to the server instance that embedded the web service and ML model}
\end{figure}

The client-server architecture \citep{Serain1995} was chosen to test the scenarios. The client module was developed\footnotemark in the programming language Python. It consisted of a script for submitting sentences to correct errors. The sentences were taken from the CoNLL-2014 database \citep{Ng2014} of 1312 sentences commonly used to evaluate solutions in GEC. The server module was composed of three software elements: 1) Nginx used as a reverse proxy \citep{Reese2008}, 2) The Flask framework \citep{Mufid2019} used to provide the web service that encapsulates GECToR as every request will be submitted to GECToR to correct grammatical errors and 3) Prometheus used to extract metrics from the environment and services \citep{Brazil2018}. Figure \ref{img:soft-arch} shows the general architecture of the simulation software used.

\begin{figure}[ht]%
\centering
\includegraphics[width=0.70\textwidth]{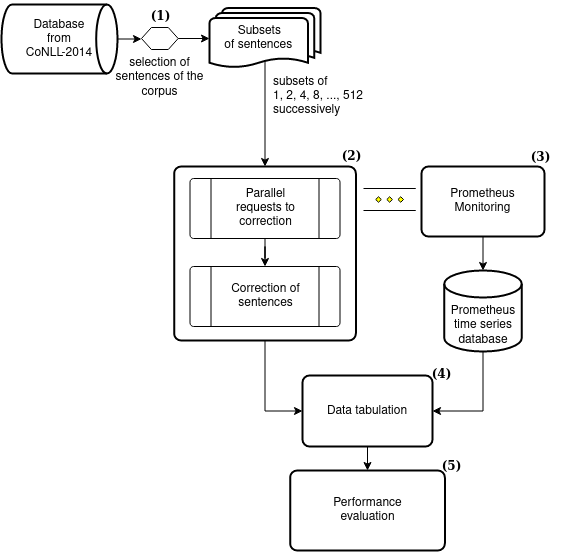}
\caption{\label{img:simulation} Simulation flow: data pre-processing to generate subsets of sentences, simulation with requests in batch, tabulation, and evaluation of data generated by the requests}
\end{figure}

\footnotetext{Source code publicly available at GitHub: infrastructure configuration, at \url{https://github.com/elaynelemos/tcc-infra}; RESTFul API for GECToR, at \url{https://github.com/elaynelemos/gector-api}; client simulation and pre-processing, at \url{https://github.com/elaynelemos/gector-client-simulator}.}

The steps performed are described in the diagram of Figure \ref{img:simulation}. Both the client and the monitoring units remained unchanged after provisioning, with the simulation per instance following the same flow described here. Before starting the simulation in each instance, (1) the selection and division of the input sentences from the test dataset were randomly performed, creating subsets equivalent to $2^N$ for $N$ ranging from 0 to 9. In the simulation, the client was in charge of sending the number of requests (2) with the sentences chosen in the pre-processing phase, and the API received the requests, correcting each sentence in a parallel and independent way. The interaction between client and server generated system and service metrics through an exporter node (3) installed in the API instances so that Prometheus could store and interpret the data, while the client stored the return of the simulation script. After the simulation, the data were tabulated (4) for performance analysis (5) of GECToR, according to the selected metrics of latency, memory utilization, and CPU utilization.

\subsection{Performance evaluation metrics}\label{sec:evaluation}

Ten experiments were carried out for each scenario describe in Table \ref{tab:instances}. Each experiment involved submitting a varying number of sentences to the server simultaneously, following a pattern of $2^N$, where $N$ ranges from 0 to 9. This resulted in sets of Number of Sentences (NS): 1, 2, 4, 8, 16, …, 512. The exponential progression was chosen to have a clear baseline for identifying performance degradation by the latency increase \citep{Saljooghinejad2015,Jiang2015,Yan2023}. Each experiment was repeated ten times to measure the statistic. The metrics selected in this study were:

\begin{itemize}
    \item real-time latency: system response time;
    \item hardware resource consumption: memory consumption and CPU or GPU time;
    \item infrastructure cost: the cost associated with provisioning cloud solutions.
\end{itemize}

\section{Results}\label{sec:results}

For all of the following outcomes, we assume an acceptable latency below two seconds, which is a standard in the industry \citep{Yu2019,Fantinuoli2022,Iyer2005}. Each cell in Tables \ref{tab:res-aws}-\ref{tab:res-azure} displays the average latency in seconds, CPU, and memory loads as a percentage (\%) for AWS, GCP, and Azure, respectively. Results from machines F and G (those having a GPU) are shaded in gray. At this point, a note is necessary for clarity and comprehension of the tables' results. For example, in Table \ref{tab:res-aws}, column C, NS 64, cell value = [3.6–42.6–52]. It should be read as follows: sixty-four sentences in parallel processed at machine C resulted in a latency of 3.6 seconds, consuming 42.6\% of CPU and 52\% of RAM. Cells containing latency values above two seconds have been highlighted in red. The results will be discussed separately by each of the tested cloud providers.

The AWS experiments (Table \ref{tab:res-aws}) demonstrate that machines F and G (equipped with GPUs) deliver superior latency performance. However, for CPU-only configurations, the amount of processor cache memory becomes a significant factor. Machine C, with its larger cache (4 GB compared to 2 GB in others), achieves the best performance, processing up to 32 sentences concurrently in under two seconds.  Table \ref{tab:res-aws} (AWS) also reveals a sharp increase in vCPU load as the number of concurrent sentences processed increases. Notably, for machines with lower processing power (A and D), the latency threshold is achieved with less than 20\% vCPU utilization. Finally, RAM usage exhibits minimal variation with increasing concurrency.

\begin{table}[ht]
\def\arraystretch{1.5}%
\fontsize{7pt}{7pt}\selectfont
\centering
\caption{\label{tab:res-aws} In order, average latency (s), vCPU, and memory loads (\%) in each instance by Number of Sentences (NS) in parallel for AWS}
\begin{tabular}{lccccccc}
\hline
\textbf{NS} & \textbf{A}    & \textbf{B}       & \textbf{C}       & \textbf{D}       & \textbf{E}       & \cellcolor[gray]{0.85}\textbf{F}       & \cellcolor[gray]{0.85}\textbf{G}       \\ \hline
1     & 1.5 - 1.5 - 84      & 0.5 - 8.1 - 63   & 0.5 - 0.5 - 60   & 1.4 - 5.1 - 86   & 0.8 - 0.8 - 65   & \cellcolor[gray]{0.85}1.2 - 8 - 87   & \cellcolor[gray]{0.85}0.3 - 0.2 - 69   \\
2     & 0.7 - 2.4 - 84      & 0.3 - 1.0 - 63   & 0.3 - 1.4 - 60   & 0.5 - 6.4 - 86   & 0.2 - 0.5 - 64   & \cellcolor[gray]{0.85}0.4 - 2.3 - 86   & \cellcolor[gray]{0.85}0.03 - 0.3 - 69   \\
4     & 1.3 - 3.9 - 84      & 0.7 - 4.0 - 62   & 0.4 - 2.1 - 59    & 0.6 - 7.1 - 85   & 0.5 - 0.9 - 64   & \cellcolor[gray]{0.85}0.2 - 2.1 - 86   & \cellcolor[gray]{0.85}0.1 - 0.4 - 69   \\
8     & \color{red}2.7 - 12.5 - 83 & 0.9 - 6.4 - 62    & 0.6 - 4.5 - 58    & 0.9 - 6.5 - 85   & 0.8 - 2.5 - 63   & \cellcolor[gray]{0.85}0.2 - 3.2 - 86    & \cellcolor[gray]{0.85}0.1 - 0.5 - 69   \\
16    & \color{red}6.5 - 38.4 - 82 & 1.8 - 17.5 - 59    & 1.2 - 17.5 - 56    & \color{red}2.2 - 12.5 - 84  & 1.6 - 6.8 - 61    & \cellcolor[gray]{0.85}0.2 - 3.8 - 86     & \cellcolor[gray]{0.85}0.1 - 0.8 - 69     \\
32    & \color{red}9.2 - 71.8 - 82 & \color{red}2.7 - 33 - 56  & 1.8 - 26 - 53   & \color{red}3.7 - 28.1 - 83  & \color{red}2.4 - 15.5 - 59  & \cellcolor[gray]{0.85}0.3 - 3.8 - 86     & \cellcolor[gray]{0.85}0.2 - 0.9 - 69     \\
64    & \color{red}22.1 - 99.1 - 84 & \color{red}4.8 - 59.4 - 54  & \color{red}3.6 - 42.6 - 52  & \color{red}7.9 - 71.4 - 84  & \color{red}4.1 - 36.5 - 56  & \cellcolor[gray]{0.85}0.5 - 5 - 86   & \cellcolor[gray]{0.85}0.4 - 2.1 - 69  \\
128   & \color{red}43.2 - 100 - 85 & \color{red}9.7 - 77.8 - 55   & \color{red}6.9 - 62.7 - 52  & \color{red}14.6 - 95.4 - 85 & \color{red}7.9 - 62.6 - 55  & \cellcolor[gray]{0.85}0.9 - 7.1 - 86    & \cellcolor[gray]{0.85}0.7 - 3.9 - 69  \\
256   & \color{red}55.1 - 100 - 86 & \color{red}17.9 - 88.5 - 55 & \color{red}13 - 85.6 - 53 & \color{red}29.5 - 99 - 86 & \color{red}14.9 - 91.2 - 55 & \cellcolor[gray]{0.85}1.6 - 14.3 - 86    & \cellcolor[gray]{0.85}1.2 - 14.4 - 69  \\
512   & \color{red}58.1 - 100 - 86 & \color{red}29.5 - 83.7 - 56  & \color{red}23.3 - 78.9 - 54 & \color{red}42.2 - 99.9 - 87 & \color{red}24.3 - 90.3 - 55 & \cellcolor[gray]{0.85}\color{red}2.9 - 34 - 86 & \cellcolor[gray]{0.85}\color{red}2.5 - 30.1 - 69 \\
\hline
\end{tabular}
\end{table}

Tests on GCP (Table \ref{tab:res-gcp}) confirm that machines with GPUs have better performance, with all results on GCP following the same trend as AWS in terms of the latency threshold. Only one exception was observed in experiment F 256 [2.4–25.5–94], which resulted in a latency of 2.4 seconds, beyond the chosen limit. A reasonable interpretation for this difference may be the existence of background variables or interactions in the experimental tests that introduce noise in the results. In a broad sense, the performance of GPU machines in GCP is similar to the results in AWS. Low processing machines (e.g., A and D) crossed the threshold around 20\% of vCPU load, and machine E crossed this limit with only 9.6\% of vCPU load. The hypothesis of a performance gain related to the processor cache memory size is corroborated by the various results for GCP, especially B 16 [1.8–12.2–4] and C 16 [1.8–12.2–4]. Machine C exhibited identical performance in terms of latency, vCPU, and memory consumption, even though it has four times less processing power (vCPU). In fact, the major difference in favor of the C machine, like the configuration on the AWS version of the experiment, resides in the cache memory. Finally, RAM consumption again does not seem to interfere with the overall performance, since experiments with latency values before or after the limit region have very close values in terms of memory consumption.

\begin{table}[ht]
\def\arraystretch{1.5}%
\fontsize{7pt}{7pt}\selectfont
\centering
\caption{\label{tab:res-gcp} In order, average latency (s), vCPU,  and memory loads (\%) in each instance by Number of Sentences (NS) in parallel for Google Cloud (GCP)}
\begin{tabular}{lccccccc}
\hline
\textbf{NS} & \textbf{A}    & \textbf{B}       & \textbf{C}       & \textbf{D}       & \textbf{E}       & \cellcolor[gray]{0.85}\textbf{F}       & \cellcolor[gray]{0.85}\textbf{G}       \\ \hline
1     & 1.6 - 0.7 - 66      & 0.3 - 0.3 - 47   & 0.3 - 0.4 - 47   & 1.2 - 0.6 - 65   & 1.2 - 0.2 - 48   & \cellcolor[gray]{0.85}1.3 - 1.8 - 94   & \cellcolor[gray]{0.85}0.2 - 0.4 - 76   \\
2     & 1.3 - 3.6 - 66      & 0.3 - 0.7 - 47   & 0.3 - 0.9 - 47   & 1.1 - 2.7 - 66   & 1.1 - 0.5 - 48   & \cellcolor[gray]{0.85}0.8 - 2.7 - 94   & \cellcolor[gray]{0.85}0.1 - 0.5 - 76   \\
4     & 1.3 - 6.7 - 66      & 1.0 - 1.7 - 47   & 1.0 - 1.6 - 47   & 0.7 - 5.7 - 66   & 0.7 - 0.9 - 48   & \cellcolor[gray]{0.85}0.5 - 4.2 - 94   & \cellcolor[gray]{0.85}0.1 - 0.6 - 76   \\
8     & \color{red}3.0 - 20.1 - 66 & 1.1 - 7.2 - 47    & 1.1 - 6.6 - 48    & 1.1 - 8 - 66   & 1.1 - 4.2 - 48   & \cellcolor[gray]{0.85}0.2 - 5.7 - 94    & \cellcolor[gray]{0.85}0.2 - 0.9 - 76   \\
16    & \color{red}6.9 - 49.2 - 67 & 1.8 - 12.2 - 47    & 1.8 - 11 - 48    & \color{red}2.5 - 19.6 - 67  & \color{red}2.5 - 9.6 - 48    & \cellcolor[gray]{0.85}0.3 - 6.7 - 94     & \cellcolor[gray]{0.85}0.3 - 1.3 - 76     \\
32    & \color{red}12.9 - 81.9 - 69 & \color{red}2.6 - 25.3 - 47  & \color{red}2.6 - 28.1 - 48   & \color{red}4.6 - 37.4 - 68  & \color{red}4.6 - 17.9 - 48  & \cellcolor[gray]{0.85}0.4 - 7.4 - 94     & \cellcolor[gray]{0.85}0.4 - 2.3 - 76     \\
64    & \color{red}25.7 - 99.2 - 71 & \color{red}5.0 - 48.9 - 48  & \color{red}5.0 - 56.1 - 49  & \color{red}8.3 - 71.9 - 69  & \color{red}8.3 - 35.5 - 49  & \cellcolor[gray]{0.85}0.8 - 8.7 - 94  & \cellcolor[gray]{0.85}0.6 - 5.2 - 76  \\
128   & \color{red}43.2 - 100 - 72 & \color{red}9.9 - 75.4 - 49   & \color{red}9.9 - 80.1 - 49  & \color{red}16.8 - 99.6 - 70 & \color{red}16.8 - 59.9 - 49  & \cellcolor[gray]{0.85}1.4 - 12.8 - 94    & \cellcolor[gray]{0.85}1.0 - 6.9 - 76  \\
256   & \color{red}55.3 - 100 - 73 & \color{red}18.6 - 93.8 - 50 & \color{red}18.6 - 99.1 - 50 & \color{red}33.2 - 100 - 71 & \color{red}33.2 - 83.4 - 50 & \cellcolor[gray]{0.85}\color{red}2.4 - 25.5 - 94    & \cellcolor[gray]{0.85}1.7 - 17.3 - 76  \\
512   & \color{red}62.3 - 100 - 73 & \color{red}39.5 - 91.9 - 50  & \color{red}39.5 - 100 - 50 & \color{red}48.1 - 100 - 72 & \color{red}48.1 - 93.3 - 51 & \cellcolor[gray]{0.85}\color{red}4.3 - 54.5 - 94 & \cellcolor[gray]{0.85}\color{red}2.9 - 29.6 - 76 \\
\hline
\end{tabular}
\end{table}

As expected for those machines with a GPU, Azure (Table \ref{tab:res-azure}) also presents better performance (F and G in all experiments). These results are in accordance with outcomes from AWS and GCP. The latency limit is exceeded, for example, when the vCPU load reaches values between 20\% and 23\% in experiments with low processing power machines, e.g., A 8 [3–19.9–68] and E 32 [2.6–22.3–49], respectively. And once again, RAM consumption does not interfere with processing response time due to similar values of memory load for both up and down latency thresholds.

\begin{table}[ht]
\def\arraystretch{1.5}%
\fontsize{7pt}{7pt}\selectfont
\centering
\caption{\label{tab:res-azure} In order, average latency (s), vCPU, and memory loads (\%) in each instance by Number of Sentences (NS) in parallel for Microsoft Azure}
\begin{tabular}{lccccccc}
\hline
\textbf{NS} & \textbf{A}    & \textbf{B}       & \textbf{C}       & \textbf{D}       & \textbf{E}       & \cellcolor[gray]{0.85}\textbf{F}       & \cellcolor[gray]{0.85}\textbf{G}       \\ \hline
1     & 0.8 - 0.5 - 67      & 0.2 - 0.3 - 49   & 0.1 - 0.5 - 50   & 0.8 - 0.8 - 74   & 0.2 - 0.4 - 48   & \cellcolor[gray]{0.85}0.2 - 0.8 - 82   & \cellcolor[gray]{0.85}0.1 - 0.5 - 41   \\
2     & 0.9 - 2.9 - 67      & 0.3 - 0.6 - 49   & 0.3 - 0.7 - 50   & 0.7 - 1.6 - 74   & 0.2 - 0.6 - 48   & \cellcolor[gray]{0.85}0.1 - 0.9 - 82   & \cellcolor[gray]{0.85}0.1 - 0.5 - 41   \\
4     & 1.2 - 5.8 - 67      & 1.5 - 2.2 - 49   & 0.5 - 1.6 - 50   & 0.7 - 4.5 - 74   & 0.7 - 1.4 - 48   & \cellcolor[gray]{0.85}0.1 - 1.0 - 82   & \cellcolor[gray]{0.85}0.1 - 0.5 - 41   \\
8     & \color{red}3.0 - 19.9 - 68 & 1.2 - 11 - 49    & 0.8 - 4.4 - 50    & 1.4 - 8.6 - 75   & 1.1 - 4.8 - 48   & \cellcolor[gray]{0.85}0.1 - 1.3 - 82    & \cellcolor[gray]{0.85}0.1 - 0.6 - 41   \\
16    & \color{red}7.1 - 55.5 - 69 & 1.7 - 17 - 49    & 1.6 - 9.6 - 51    & \color{red}2.7 - 21.7 - 76  & 1.7 - 10.5 - 48    & \cellcolor[gray]{0.85}0.2 - 1.8 - 82     & \cellcolor[gray]{0.85}0.2 - 0.9 - 41     \\
32    & \color{red}12.2 - 81 - 70 & \color{red}2.6 - 29.7 - 50  & \color{red}2.6 - 22.4 - 51   & \color{red}5.3 - 46 - 78  & \color{red}2.6 - 22.3 - 49  & \cellcolor[gray]{0.85}0.3 - 2.8 - 82     & \cellcolor[gray]{0.85}0.3 - 1.3 - 41     \\
64    & \color{red}23.2 - 98.1 - 72 & \color{red}4.8 - 51.8 - 50  & \color{red}5.0 - 52.4 - 53  & \color{red}9.6 - 72.7 - 80  & \color{red}4.9 - 46.9 - 51  & \cellcolor[gray]{0.85}0.5 - 5.4 - 82   & \cellcolor[gray]{0.85}0.5 - 2.7 - 41  \\
128   & \color{red}42.5 - 100 - 73 & \color{red}9.4 - 74.6 - 51   & \color{red}9.8 - 78.1 - 54  & \color{red}20 - 95.9 - 81 & \color{red}9.6 - 75.8 - 52  & \cellcolor[gray]{0.85}0.8 - 8.6 - 82    & \cellcolor[gray]{0.85}0.8 - 5.5 - 41  \\
256   & \color{red}54.5 - 100 - 74 & \color{red}17.9 - 92.1 - 52 & \color{red}18.6 - 98.8 - 55 & \color{red}37.8 - 100 - 82 & \color{red}18.2 - 98.6 - 53 & \cellcolor[gray]{0.85}1.5 - 16.7 - 82    & \cellcolor[gray]{0.85}1.4 - 10.7 - 41  \\
512   & \color{red}59.1 - 100 - 75 & \color{red}39.2 - 89.8 - 52  & \color{red}38.6 - 99.5 - 56 & \color{red}52.2 - 100 - 83 & \color{red}36.7 - 98 - 54 & \cellcolor[gray]{0.85}\color{red}2.7 - 34.9 - 82 & \cellcolor[gray]{0.85}\color{red}2.5 - 24.9 - 41 \\
\hline
\end{tabular}
\end{table}

From the results, AWS, GCP, and Azure showed similar behavior, even though there are uncontrolled background variables that might have an impact on the experiments, such as differences in the underlying OS and its dependencies, different processors even for the same instance types, noisy neighbors in the data centers, and resource consumption in shared servers. Regarding latency, the trend was repeated in almost all experimental arrangements. The response time trend, vCPU load, and memory consumption were preserved in the remaining configurations.

Revisiting the research questions posed in Subsection \ref{sec:questions}, we can now provide answers based on the comprehensive analysis presented in this section.

\textbf{How much RAM is required for acceptable performance?}  It may be defined taking into account the size of the model whose file contains the network weights. As it resides in RAM, only one GB more is needed for the OS and the software composing the web service. The results showed no increase in memory consumption despite the increase in the concurrent messages processed. Thus, for this NLP application, there is evidence that the amount of RAM did not interfere with latency since the average memory consumption between acceptable and unacceptable latencies was similar. In this way, it indicates that acquiring more memory is a cost that may be avoided since it is an expensive resource. For example, machine G has twice as much memory as machine F, but they presented very close average latency values for the same scenarios. However, the last one (machine G) cost 43\%, 35\%, and 43\% more, for AWS, GCP, and Azure, respectively (Table \ref{tab:cost}).

\textbf{How many virtual CPUs (vCPUs) are necessary to handle the expected workload?} The results showed that more vCPU power reduces the solution’s latency. However, they also indicated that the amount of cache memory is a relevant parameter for architectures without GPUs. This difference can be explained because the processor cache memory is of type SRAM, which is ten times faster than main memory (DRAM). So, the processor gets the data from the cache memory faster \citep{Chi2016,Hameed2013,Wang2008}. This result is important, as it indicates a cost reduction of around 50\% for machine C compared to machine E, as seen in Table \ref{tab:cost}. It is essential to highlight that machine C achieved the best overall performance among those with constrained resources. Still on these machines, results show that the acceptable latency threshold was often exceeded even with little vCPU load (around 20\%). An alternative to bypass this limitation would be to create a queue in the application layer to control submission flow, taking this processing threshold into account.

\textbf{Can the model achieve acceptable performance without a GPU?} GPU solutions obtained the best results, as expected. However, they are, on average, 300\% more expensive, as can be verified by Table \ref{tab:cost}. Thus, for a feasibility test of a product such as a POC, the results show that non-GPU alternatives are viable.

\begin{table}[ht]
\def\arraystretch{1.5}%
\fontsize{7pt}{7pt}\selectfont
\centering
\caption{\label{tab:cost} Experiment instances monthly equivalent cost rates in US\$  by cloud provider}
\begin{tabular}{lccccccc}
\hline
\textbf{Cloud} & \textbf{A} & \textbf{B}   & \textbf{C} & \textbf{D} & \textbf{E}  & \textbf{F}   & \textbf{G}   \\ \hline
AWS & 110.16 & 220.32 & 133.63 & 164.16 & 260.64 & 378.72 & 541.44 \\
GCP & 100.44 & 200.87 & 230.89 & 124.10 & 248.21 & 388.80 & 525.60 \\
Azure & 95.76 & 191.52 & 276.48 & 121.68 & 243.36 & 383.98 & 548.96 \\ \hline
\end{tabular}
\end{table}

Revisiting the research questions posed in Subsection \ref{sec:questions}, we can now provide answers based on the comprehensive analysis presented in this section.

\section{Conclusion}\label{sec:conclusion}

This study evaluated the feasibility of conducting POCs for DL models in cloud environments with limited computational resources. Using the GECToR model for GEC, we compared deployments across AWS, GCP, and Azure, analyzing latency, CPU and GPU utilization, and associated costs.

The findings reveal three main insights. First, RAM requirements can be defined primarily by the size of the pre-trained model’s weight file, since memory consumption remained stable across all concurrency levels. Second, CPU performance significantly affects latency, and the results highlight cache memory size as a decisive factor in non-GPU setups. Machines with larger cache achieved up to 50\% cost reduction compared to higher-tier configurations. Finally, GPU instances offered superior performance but at an average cost approximately 300\% higher.

Therefore, this research indicates that deploying DL models for experimental and POC applications may be feasible and cost-effective without GPUs, especially when processor cache memory is optimized. These results are valuable for startups and researchers seeking low-cost alternatives for prototyping DL solutions in the cloud.

\section{Recommendation and limitations}\label{sec:future}

This research has two main limitations. First, it assessed a single NLP model (GECToR), which limits generalization across other DL domains. Future work should extend the analysis to models from Computer Vision and other fields, where data dimensionality and memory requirements differ substantially. Such comparisons could deepen the understanding of how architecture and input size affect cost and latency in cloud deployments. Second, external and uncontrolled cloud factors such as shared-resource contention, underlying hardware variations, and OS differences may have influenced results. Controlling or statistically modeling these sources of variability would increase experimental robustness.

Future studies should also incorporate network latency analysis, API and software optimizations, and dynamic pricing models, including reserved or spot instances, as well as regional and smaller providers that may offer more competitive options. Addressing these points will expand the current findings and provide a broader understanding of resource-efficient DL deployment strategies across diverse platforms and model types.

\bibliography{references}
\end{document}